\documentclass[letterpaper]{jpconf}
\usepackage[utf8]{inputenc}
\usepackage{amsmath}
\usepackage{amsfonts}
\usepackage{amssymb}
\usepackage{graphicx}
\usepackage{cite}
\usepackage{slashed}
\usepackage{array}
\usepackage{mathrsfs}
\usepackage{transparent}

\newcommand{\Psib}{\bar{\Psi}}

\newcommand{\Ds}{\slashed{D}}

\newcommand{\A}{\textrm{A}}
\newcommand{\F}{\textrm{F}}
\newcommand{\drm}{\textrm{d}}
\newcommand{\Det}{\textrm{Det}}
\newcommand{\Zz}{\mathcal{Z}}
\newcommand{\arm}{\textrm{a}}
\newcommand{\av}[1]{\big\langle\!\big\langle  #1 \big\rangle\!\big\rangle}

\newcommand{\totint}{\int_{-\infty}^{\infty}}
\newcommand{\half}{\frac{1}{2}}

\begin{document}

\title{Plane Wave Backgrounds in the Worldline Formalism}

\author{\underline{James P. Edwards}
$^{1}$
 and Christian Schubert
$^{1}$
}
\address{$^{1}$Instituto de Física y Matemáticas, Universidad Michoacana de San Nicolás de Hidalgo.}

\ead{jpedwards@cantab.net, christianschubert137@gmail.com}

\begin{abstract}
Plane-wave backgrounds play a special role in strong-field QED as examples of a non-trivial field configuration that remains simple enough to be treated analytically whilst still leading to rich physical consequences. Although great progress has been made applying standard field theory techniques to QED in plane wave backgrounds, the calculations tend to be quite long and complicated. Yet, both in vacuum and in constant backgrounds, the first quantised, string-inspired ``Worldline Approach'' to field theory has a long history of offering substantial simplifications and calculational efficiency.

We present a new, general approach to incorporating plane wave backgrounds into the Worldline Formalism that extends initial work using a semi-classical approach by Ilderton and Torgrimsson (who also participated in LPHYS'21). The method uses resummation techniques to take the background into account non-perturbatively and yields ``Master Formulae'' for the effective action and scattering amplitudes in the background. It is hoped that this may offer an alternative tool to studying QED in plane waves that may streamline otherwise complex calculations, as has been achieved in the better explored constant field case.
\end{abstract}

\section{Introduction}
\label{secIntro}
The behaviour of quantum fields in the presence of electromagnetic backgrounds -- such as those produced at high intensity laser experiments -- can be very different to that of fields in vacuum. In particular, additional interactions with photons from the background make possible various new phenomena, some of which form the basis for current and future experimental searches. It is important that theoretical calculations take into account these effects with realistic models of such external fields, especially considering the high precision of experiments planned to come online in the years ahead. Moreover, the mathematical structure of the theory in a background field is of theoretical interest, especially its strong field asymptotics in the context of the famous Ritus-Narozhny conjecture \cite{Ritus, Narozhnyi:1980dc} regarding the perturbative expansion parameter of dominant loop contributions that first arose for constant-crossed backgrounds.

In this contribution we shall consider one-loop $N$-photon amplitudes in three cases: quantum electrodynamics (QED) in vacuum, in constant backgrounds and in an arbitrary \textit{plane wave} background, using the \textit{Worldline Formalism} of quantum field theory (see \cite{Strass1, ChrisRev, 103, UsRep}). This alternative approach to field theory, appreciated for its computational advantages and efficient representation of such scattering amplitudes encoded in ``Master Formulae,'' has only recently been adapted to a general treatment in plane wave fields in \cite{NPlane} building on previous work focussing on the on-shell case in \cite{AntonPlane} which treated a specific $N=2$-photon process. 

These backgrounds (described in more detail below) are of both phenomenological importance as models of the fields produced by the intense lasers mentioned above, and of intrinsic mathematical interest as fields for which the Dirac equation can be solved exactly \cite{Gies, Gitman} (as explained later, this allows field theory processes to be studied non-perturbatively with respect to the background field, albeit in perturbation theory in the fine structure constant, $\alpha \equiv \frac{e^{2}}{4\pi}$ with $e$ the electric charge, according to the Furry picture \cite{Furry, Seipt:2017ckc}). As such they are very special examples that permit analytic treatment without losing most of the essential features of QED in more general backgrounds that cannot be solved exactly. 

On the other hand, despite great efforts towards studying various processes in these backgrounds, calculations in the standard approach often turn out to be rather laborious except for the simplest cases (see, for example, \cite{ Schwinger:1951nm, Nikishov:1964zza, Nikishov:1964zz, Baier:1967zzc, Batalin:1970it, Adler:1971wn, Neville:1971uc, Tsai:1974df, Baier:1974qq, Becker:1974en, DeRaad:1974mwa,  Urrutia:1977xb, Morozov:1981pw, DiPiazza:2007yx, Bragin:2020akq, DiPiazza:2020kze}). This work extending the Worldline Formalism to plane wave backgrounds is motivated by its success in providing compact ``Master Formulae'' for scattering amplitudes in vacuum and constant backgrounds, that have led to useful insight and efficient analysis of field theory processes and their gauge structure in those cases; here we describe these results and the derivation of a Master Formula for one-loop $N$-photon amplitudes in a plane wave background recently presented for the first time in \cite{NPlane}.

We begin in section \ref{secVac} with an introductory description of the worldline approach to quantum field theory, illustrating its application in vacuum, before recapitulating the worldline description of QED in a constant background in section \ref{secConst}. In section \ref{secPlane} we show how a plane wave background can be treated in the Worldline Formalism before concluding. Contrary to the majority of the works cited, we present our results in Minkowski space throughout (recalling that functional integration is more conveniently defined after Wick rotation to Euclidean space).

\section{Photon amplitudes in vacuum}
\label{secVac}
To set the scene, we consider photon amplitudes in QED, starting from the Dirac action coupling the electron field, $\Psi(x)$, to the Maxwell field with gauge potential $\textrm{A}(x) = A_{\mu}(x)dx^{\mu}$,
\begin{equation}
	S[\Psi, A] = \int d^{D}x\big[ -\!\frac{1}{4}\tr(F^{2}) + \Psib(i \Ds - m)\Psi \big]\,,
	\label{eqSPsiA}
\end{equation}
with covariant derivative $D_{\mu} := \partial_{\mu} + ieA_{\mu}$ and field strength tensor $\F(x) = \drm\A(x)$. Here we consider $\A$ to be a semi-classical field, consisting of a ``background field'' and a ``photon field'' so that $\A(x) = \bar{\A}(x) + \A^{\gamma}(x)$. For general backgrounds, $\bar{\A}(x)$, a full analytic solution to the system described by (\ref{eqSPsiA}) is out of reach, but amongst the few background fields that allow for an exact classical solution are the constant and plane wave backgrounds we consider in later sections.

\subsection{Worldline representation --- effective action}
For now we consider photon scattering in vacuum, for which we set $\bar{\A}(x) = 0$, which we  access via the one-loop effective action for the photon field, $\A^{\gamma}(x)$. As outlined in \cite{Strass1} and reviewed in \cite{UsRep, ChrisRev}, this is obtained by integrating out the fermion degrees of freedom according to
\begin{align}
	\e^{i\Gamma[A]} &:= \int \mathscr{D}\Psib(x)\mathscr{D}\Psi(x) \exp{\big[i \int d^{D}x\, \Psib(x)(i \gamma \cdot D- m)\Psi(x) \big]}\nonumber \\
	&= i\textrm{Det}\big[ i\gamma \cdot D - m \big] \nonumber \\
	&= \frac{i}{2}\textrm{Det}^{\frac{1}{2}}\big[ (\gamma \cdot D)^{2} + m^{2} \big]\,.
	\label{eqGammaDef}
\end{align}
A \textit{first quantised} representation of the functional determinant is obtained by first extending a familiar trick to operators, $\mathcal{O}$, namely that $\log \Det[\mathcal{O}] = \Tr \log[\mathcal{O}]$; then employing Schwinger's representation of the logarithm and evaluating the trace in the position basis we arrive at
\begin{align}
	\Gamma[A] &= -\frac{1}{2}\int_{0}^{\infty} \frac{dT}{T}\e^{-im^{2}T} \int d^{D}x \, \big\langle x \big| \e^{-iT(\gamma \cdot D)^{2} } \big| x \big\rangle \nonumber\\
	&=-\frac{1}{2}\int_{0}^{\infty} \frac{dT}{T}\e^{-im^{2}T} \oint_{PBC}\mathscr{D}x(\tau) \oint_{ABC}\mathscr{D}\psi(\tau) \,\e^{iS[x, \psi]}\,,
	\label{eqGammaWL}
\end{align}
where the last line uses the path integral representation of the (diagonal) matrix elements, summing over $T$-periodic trajectories, $x^{\mu}(\tau)$, that represent the orbital degrees of freedom of the field and anti-periodic Grassmann fields, $\psi(\tau)$, on these worldlines that produce the spin interaction with the background field. The ``worldline action''  inherited from the Dirac operator (that has been transformed to second order form \cite{PhysRev.109.193, Hostler:1985vb} in the first line of (\ref{eqGammaWL})) is given by
\begin{equation}
	S[x, \psi] = \int_{0}^{T}d\tau \Big[ \frac{\dot{x}^{2}}{4} + \frac{i}{2}\psi \cdot \dot{\psi} - eA(x)\cdot \dot{x} + ie\psi \cdot F(x) \cdot \psi \Big]\,,
	\label{eqSWL}
\end{equation}
describing dynamics of a relativistic spin-$1/2$ point particle minimally coupled to the background ($\tau$ is the particle's proper time and $T$ the proper time of propagation around the loop).

In scalar QED the effective action follows from integrating out the spin-zero Klein-Gordon field and leads to an analogous formula differing only by the absence of the Grassmann fields and overall normalisation (for degrees of freedom and the difference in statistics)
\begin{align}
	\Gamma_{\textrm{scal}}[A] = \int_{0}^{\infty} \frac{dT}{T}\e^{-im^{2}T} \oint_{PBC}\mathscr{D}x(\tau)\,  \e^{i\int_{0}^{T} d\tau \big[ \frac{\dot{x}^{2}}{4} - eA(x)\cdot \dot{x}  \big]}\,.
	\label{eqGammaScalWL}
\end{align}
These proper time integral representations of the effective action are the basic objects of the Worldline Formalism. It is worth pointing out that analogous representation for the field theory propagators in an electromagnetic background were first suggested by Feynman in the appendices of \cite{Feynman:1950ir, Feynman:1951gn} (recently extended to a complete worldline description in vacuum in \cite{fppaper1, fppaper2}).

Although less well-known than the standard formalism, the worldline approach has proven to offer various conceptual and computational advantages. These largely stem from the fact that, as in string theory, it combines multiple Feynman diagrams into single, compact integral representations, thereby manifesting symmetry under exchange of external legs. As such it offers superior organisation of gauge information and removal of spurious divergences that would normally cancel between diagrams already at the level of the integrand. Moreover, as in the calculations presented below virtual loop momenta are already integrated out. 

\subsection{Scattering amplitudes --- Master Formulae}
To obtain photon amplitudes, we now specialise $\A(x) = \A^{\gamma}(x)$ to represent asymptotic wavefunctions of $N$ photons with fixed polarisations and momenta by
\begin{equation}
	A_{\mu}^{\gamma}(x) = \sum_{i = 1}^{N}\varepsilon_{i\mu}\,\e^{i k_{i}\cdot x}
	\label{eqAGamma}
\end{equation}
and expand the interactions in (\ref{eqGammaWL}), (\ref{eqGammaScalWL}) to multi-linear order in the $\varepsilon_{i}$. Doing this represents each photon emission or absorption off the loop by a ``vertex operator'' familiar from string theory,
\begin{align}
	V^{\gamma}[k, \varepsilon] &:= \int_{0}^{T}d\tau \big[\varepsilon\cdot \dot{x}(\tau) + \psi(\tau) \cdot f \cdot \psi(\tau)\big]\e^{i k \cdot x(\tau)} \,, \nonumber\\
	V^{\gamma}_{\textrm{scal}}[k, \varepsilon] &:= \int_{0}^{T}d\tau \, \varepsilon\cdot \dot{x}(\tau)\, \e^{i k \cdot x(\tau)} \,,
	\label{eqVertices}
\end{align}
where in the first line $f_{\mu\nu} := k_{\mu}\varepsilon_{\nu} - \varepsilon_{\mu}k_{\nu}$ is the (linearised) photon field strength tensor. This procedure leads easily to the following representations of the $N$-photon amplitudes (first arising from Bern and Kosower's analysis of the infinite tension limit of string theory amplitudes \cite{Bern:1990ux, Bern:1991aq}):
\begin{align}
	\hspace{-2em}\Gamma[k_{1}, \varepsilon_{1};\ldots;k_{N}, \varepsilon_{N}] &= -\frac{1}{2}(-ie)^{N}\int_{0}^{\infty}\frac{dT}{T}\e^{-im^{2}T} \oint_{PBC}\mathscr{D}x(\tau)\,\e^{i\int_{0}^{T}d\tau \,\frac{\dot{x}^{2}}{4}} \oint_{ABC}\mathscr{D}\psi(\tau)\, \e^{i\int_{0}^{T}d\tau \,\frac{i}{2}\psi \cdot \dot{\psi}} \, \prod_{i=1}^{N} V^{\gamma}[k_{i}, \varepsilon_{i}] \nonumber \\
	\hspace{-2em}\Gamma_{\textrm{scal}}[k_{1}, \varepsilon_{1};\ldots;k_{N}, \varepsilon_{N}] &= (-ie)^{N}\int_{0}^{\infty}\frac{dT}{T}\e^{-im^{2}T} \oint_{PBC}\mathscr{D}x(\tau)\,\e^{i\int_{0}^{T}d\tau \,\frac{\dot{x}^{2}}{4}}\, \prod_{i=1}^{N} V^{\gamma}_{\textrm{scal}}[k_{i}, \varepsilon_{i}] \,.
	\label{eqGammasVertex}
\end{align}
Note that both integrals are now Gaussian in their corresponding variables.

To compute these path integrals, then, we only require the Green functions of the free kinetic operator and overall normalisation. In the orbital sector, for closed loops this operator has a zero mode, $x_{0} = \textrm{const}$, so we separate this by expanding about the loop centre of mass, 
\begin{equation}
	\hspace{-2em}x^{\mu}(\tau) = x_{0}^{\mu} + q^{\mu}(\tau)\,,  \qquad x_{0}^{\mu} = \frac{1}{T}\int_{0}^{T}d\tau \, x^{\mu}(\tau)\,, \qquad q^{\mu}(0) = 0 = q^{\mu}(T)\,, \qquad \int_{0}^{T}d\tau\, q^{\mu}(\tau) = 0\,,
	\label{eqDefxq}
\end{equation}
which has led to so-called ``string-inspired'' (SI) boundary conditions on the deviation $q(\tau)$ which remains to be integrated over: $\oint \mathscr{D}x(\tau) \longrightarrow \int d^{D}x_{0} \int_{\textrm{SI}} \mathscr{D}q(\tau)$. With this, the appropriate Green functions in the Hilbert space orthogonal to the zero mode are defined by
\begin{align}
	\big\langle q^{\mu}(\tau_{i})q^{\nu}(\tau_{j})\big\rangle_{\perp} &= -G_{Bij}\eta^{\mu\nu} \,, \qquad G_{Bij} \equiv G_{B}(\tau_{i}, \tau_{j}) = |\tau_{i} - \tau_{j}| - \frac{(\tau_{i} - \tau_{j})^{2}}{T}\,, \\
	\big\langle \psi^{\mu}(\tau_{i})\psi^{\nu}(\tau_{j})\big\rangle &= \frac{1}{2}G_{Fij}\eta^{\mu\nu} \,, \qquad G_{Fij} \equiv G_{F}(\tau_{i}, \tau_{j}) = \sigma(\tau_{i} - \tau_{j}) \,.
	\label{eqGreenFns}
\end{align}
In the scalar case, we can borrow another trick from string theory by exponentiating the prefactor in the vertex operator, writing $V^{\gamma}_{\textrm{scal}}[k, \varepsilon] = \int_{0}^{T}d\tau \, \e^{i k \cdot x(\tau) + \varepsilon\cdot \dot{x}(\tau)}\Big|_{\textrm{lin }\varepsilon}$ which allows the path integral to be computed by completing the square. With the free particle path integral normalisation $\int_{\textrm{SI}}\mathscr{D}q(\tau) \, \e^{i\int_{0}^{T}d\tau \, \frac{\dot{q}^{2}}{4}} = (4\pi i T)^{-\frac{D}{2}}$ we arrive at the Bern-Kosower Master formula
\begin{align}
\hspace{-1em}	\Gamma_{\textrm{scal}}[k_{1}, \varepsilon_{1};\ldots;k_{N}, \varepsilon_{N}] = (-ie)^{N}&(2\pi)^{D}\delta^{D}\big(\sum_{i = 1}^{N}k_{i}\big)\int_{0}^{\infty}\frac{dT}{T}(4\pi i T)^{-\frac{D}{2}}\e^{-im^{2}T} \nonumber \\ 
\hspace{-1em}	\prod_{i=1}^{N} &\int_{0}^{T}d\tau_{i}\, \e^{\frac{i}{2}\sum_{i, j = 1}^{N} \big[k_{i} \cdot k_{j}G_{Bij} - 2i\varepsilon_{i}\cdot k_{j}\dot{G}_{Bij} + \varepsilon_{i}\cdot \varepsilon_{j}\ddot{G}_{Bij} \big]}\Big|_{\textrm{lin }\varepsilon_{1}\ldots  \varepsilon_{N}}\,,
	\label{eqMFScalar}
\end{align}
where we note that the momentum conserving $(2\pi)^{D}\delta^{D}\big(\sum_{i = 1}^{N}k_{i}\big)$ came from integrating over the loop centre of mass (interpreted as saying that the scattering photons can neither impart nor receive momentum from the loop).

In the spinor case a similar master can be derived by using a \textit{superspace formalism} that linearises the spin part of the interaction vertex. An alternative, often more useful method decomposes contributions to the effective action according to the number of spin and orbital interactions selected from the product of vertex operators in the first line of (\ref{eqGammasVertex}):
\begin{equation}
	\Gamma = \sum_{S = 0}^{N} \Gamma_{NS}\,, \qquad \Gamma_{NS} = \sum_{\{i_{1}\ldots i_{S}\}} \Gamma_{NS}^{\{i_{1}\ldots i_{S}\}}\,,
	\label{eqSpinOrbit}
\end{equation}
in which $S$ indicates the number of times the spin part of the vertex is chosen and the sum in the second equality is over partitions of the $N$ indices to assign $S$ photons to this interaction. If we define polynomials $P_{N}$ according to the expansion of the Bern-Kosower exponent in (\ref{eqMFScalar}) by
\begin{equation}
	\e^{\frac{i}{2}\sum_{i, j = 1}^{N} \big[k_{i} \cdot k_{j}G_{Bij} - 2i\varepsilon_{i}\cdot k_{j}\dot{G}_{Bij} + \varepsilon_{i}\cdot \varepsilon_{j}\ddot{G}_{Bij} \big]}\Big|_{\textrm{lin }\varepsilon_{1}\ldots  \varepsilon_{N}} := (-i)^{N}P_{N}(\dot{G}_{ij}, \ddot{G}_{ij})\,\e^{\frac{i}{2}\sum_{i, j = 1}^{N} k_{i} \cdot k_{j}G_{Bij}}\,,
	\label{eqPN}
\end{equation}
then we can write (\ref{eqMFScalar}) more explicitly as (we omit the momentum conserving $\delta$-function)
\begin{equation}
\hspace{-1em}	\Gamma_{\textrm{scal}}[k_{1}, \varepsilon_{1};\ldots;k_{N}, \varepsilon_{N}] = (-e)^{N}\int_{0}^{\infty}\frac{dT}{T}(4\pi i T)^{-\frac{D}{2}}\e^{-im^{2}T} \prod_{i=1}^{N} \int_{0}^{T}d\tau_{i}\, P_{N}(\dot{G}_{ij}, \ddot{G}_{ij})\,\e^{\frac{i}{2}\sum_{i, j = 1}^{N} k_{i} \cdot k_{j}G_{Bij}}\,.
	\label{eqGammaScalPN}
\end{equation}
This in turn opens two avenues for the spinor case. Firstly the polynomials can be generalised for use in the spin-orbit decomposition by removing the $S$ photons assigned to the spin interaction:
\begin{equation}
\hspace{-3em}	\e^{\frac{i}{2}\sum_{i, j = 1}^{N} \big[k_{i} \cdot k_{j}G_{Bij} - 2i\varepsilon_{i}\cdot k_{j}\dot{G}_{Bij} + \varepsilon_{i}\cdot \varepsilon_{j}\ddot{G}_{Bij} \big]}\Big \vert^{\varepsilon_{i_1}=0= \cdots 
 = 0=\varepsilon_{i_S}}_{\varepsilon_{i_{S+1}}\cdots\varepsilon_{i_N}}  := (-i)^{N}P_{NS}^{\{i_{1}\ldots\,i_{S}\}}(\dot{G}_{ij}, \ddot{G}_{ij})\,\e^{\frac{i}{2}\sum_{i, j = 1}^{N} k_{i} \cdot k_{j}G_{Bij}}\,.
	\label{eqPNS}
\end{equation}
Then, along with the appropriate result for the evaluation of the spin-terms,
\begin{align}
	\label{eqWS}
	\hspace{-1.5em}W(f_{i_{1}}, \ldots f_{i_{S}}) &:= \big\langle \psi_{i_{1}}\cdot f_{i_{1}} \cdot \psi_{i_{1}} \cdots \psi_{i_{S}}\cdot f_{i_{S}} \cdot \psi_{i_{S}} \big\rangle \\
	\hspace{-1.5em}&= \sum_{\textrm{Partitions of S}}(-1)^{cy}G_{F}(i_{1}\ldots i_{n_{1}})G_{F}(i_{n_{1} + 1}i_{n_{1} + n_{2}})\ldots G_{F}(i_{n_{1} + \ldots + n_{cy-1}+1}\ldots i_{n_{1} + \ldots + n_{cy}})\,,\nonumber
\end{align}
where the sum is over inequivalent partitions of the indices between $cy$ bi-cycles defined for either the bosonic or fermionic Green function by ${G(i_{1}\ldots i_{n}):= G_{i_{1}i_{2}}G_{i_{2}i_{3}}\cdots G_{i_{n}i_{1}} Z_{n}(i_{1}\ldots i_{n})}$ with $Z_{n}(i_{1}\ldots i_{n}) = 2^{-\delta_{n2}}\tr\big(\prod_{j=1}^{n} f_{i_{j}}\big)$, we can write
\begin{equation}
	\hspace{-2em}\Gamma_{NS}^{\{i_{1}\ldots i_{S}\}} = -2(-e)^{N}\int_{0}^{\infty}\frac{dT}{T}(4\pi i T)^{-\frac{D}{2}}\e^{-im^{2}T}\prod_{i = 1}^{N}\int_{0}^{T}d\tau_{i}\, W(f_{i_{1}}, \ldots f_{i_{S}})P_{NS}^{\{i_{1}\ldots\,i_{S}\}}(\dot{G}_{ij}, \ddot{G}_{ij})\,\e^{\frac{i}{2}\sum_{i, j = 1}^{N} k_{i} \cdot k_{j}G_{Bij}}\,.
	\label{eqGammaNS}
\end{equation}
These functions completely describe the $N$-photon amplitudes in vacuum: this formalism  shall be generalised to a constant electromagnetic field and plane wave background in what follows. The second option is to follow Bern and Kosower's integration by parts procedure and apply a ``cyclic replacement rule'' to convert the scalar amplitudes into their spinor counterparts \cite{ChrisRev}.

\section{Constant electromagnetic background} 
\label{secConst}
The worldline formalism has had success in constant backgrounds, which of course are relevant for processes where the field can be approximated as slowly varying (e.g. where we may apply the LCFA). Guided by standard techniques, which incorporate the field into the Feynman rules using the exact solution to the Dirac equation, it turns out that little modification is needed to treat the field non-perturbatively in the worldline approach \cite{Reuter:1996zm, Schmidt:1993rk, Gusynin:1998bt, Shaisultanov:1995tm, Kors:1998ew, Ahmad:2016vvw}.

To facilitate the worldline computation in this background it is convenient to use Fock-Schwinger gauge centred at the loop centre of mass whereby (for a constant field)
\begin{equation}
	\bar{A}_{\mu}(x(\tau)) = -\frac{1}{2}\bar{F}_{\mu\nu}\big(x(\tau) - x_{0}\big)^{\nu} =  -\frac{1}{2}\bar{F}_{\mu\nu}q(\tau)^{\nu}\,,
\end{equation}
with which the background dependent part of the worldline action, (\ref{eqSWL}), becomes
\begin{equation}
	S_{\bar{A}}[x, \psi | \bar{F}] = \int_{0}^{T}d\tau \Big[\frac{1}{4} x\cdot \Big(-\frac{d^{2}}{d\tau^{2}} - 2e\bar{F} \partial_{\tau}\Big)\cdot x +\frac{i}{2} \psi\cdot \Big(\frac{d}{d\tau} + 2e\bar{F} \Big)\cdot \psi \Big]\,.
	\label{eqSWLF}
\end{equation}
Now the advantage of Fock-Schwinger gauge is clear: the worldline action remains quadratic in the path integration variables so the field can be accounted for exactly by modifying the kinetic terms as indicated in (\ref{eqSWLF}). Indeed, the presence of external photons (i.e. the inclusion of $\A^{\gamma}$ as in (\ref{eqAGamma})) does not spoil this Gaussian nature so the only modification required for evaluation of the effective actions are the path integral normalisations in the presence of the background:
\begin{align}
\hspace{-2em}	\underset{\textrm{SI}}{\textrm{Det}}\Big[-\partial^{2}_{\tau} - 2e\bar{F}\partial_{\tau}\Big] = (4\pi i T)^{-\frac{D}{2}} \det{}^{-\frac{1}{2}}\Big[ \frac{\sinh\Zz}{\Zz} \Big] \,, \quad \underset{\textrm{ABC}}{\textrm{Det}}\Big[\partial_{\tau} +2e\bar{F}\Big] = 2^{\frac{D}{2}} \det{}^{-\frac{1}{2}}\Big[ \cosh \Zz \Big]\,,
\end{align} 
where $\Zz := e\bar{F}T$, and the appropriate Green functions (inverses of the kinetic operators)
\begin{equation}
\hspace{-1em}	G_{Bij}\eta^{\mu\nu} \rightarrow \mathcal{G}_{Bij}^{\mu\nu}= \frac{iT}{2\Zz^{2}}\left(\frac{\Zz}{\sinh \Zz}e^{-\Zz \dot{G}_{Bij}} + \Zz \dot{G}_{Bij} - 1\right)\,, \qquad G_{Fij}\eta^{\mu\nu} \rightarrow	\mathcal{G}_{Fij}^{\mu\nu} = G_{Fij}\frac{e^{- \Zz \dot{G}_{Bij}}}{\cosh \Zz}\,,
\end{equation}
and their derivatives. This treats the external field non-perturbatively in the path integral. 

\subsection{Euler-Heisenberg Lagrangians}
We can illustrate this formalism with the $N=0$, bare path integral which recovers the (unrenormalised) one-loop Euler-Heisenberg (EH) and Weisskopf (WK) Lagrangians written in a compact form:
\begin{align}
	\mathcal{L}_{\textrm{EH}}^{(1)} &= -2\int_{0}^{\infty}\frac{dT}{T}(4\pi i T)^{-\frac{D}{2} }\e^{-im^{2} T}\det{}^{-\frac{1}{2}}\Big[ \frac{\tanh\Zz}{\Zz} \Big]\,,\\
	\mathcal{L}_{\textrm{WK}}^{(1)} &= \int_{0}^{\infty}\frac{dT}{T}(4\pi i T)^{-\frac{D}{2}} \e^{-im^{2} T}\det{}^{-\frac{1}{2}}\Big[ \frac{\sinh\Zz}{\Zz} \Big]\,.
\end{align} 
One may write the determinants in terms of the two standard secular invariants of the Maxwell field to show equivalence with the Schwinger proper-time representation these Lagrangians. For worldline calculations of these Lagrangians at higher loop order see, for example, \cite{ Reuter:1996zm, Kors:1998ew, Fliegner:1997ra, Dunne:2002qf, Dunne:2002qg, Huet:2010nt}.

\subsection{$N$-photon amplitudes}
The function forms of the master formula (\ref{eqMFScalar}) and the spin-orbit decomposition representation of the one-loop $N$-photon amplitudes (\ref{eqGammaNS}) remain valid in a constant background up to the replacements of normalisation and Green functions. For instance, for scalar QED we find \cite{Reuter:1996zm, Shaisultanov:1995tm}
\begin{align}
\hspace{-1em}	\Gamma_{\textrm{scal}}[k_{1}, \varepsilon_{1};\ldots;k_{N}, \varepsilon_{N}] = (-ie)^{N}&(2\pi)^{D}\delta^{D}\big(\sum_{i = 1}^{N}k_{i}\big)\int_{0}^{\infty}\frac{dT}{T}(4\pi i T)^{-\frac{D}{2}}\det{}^{-\frac{1}{2}}\Big[ \frac{\sinh\Zz}{\Zz} \Big]\e^{-im^{2}T} \nonumber \\ 
\hspace{-1em}	\prod_{i=1}^{N} &\int_{0}^{T}d\tau_{i}\, \e^{\frac{i}{2}\sum_{i, j = 1}^{N} \big[k_{i} \cdot \mathcal{G}_{Bij}\cdot k_{j} - 2i\varepsilon_{i}\cdot \dot{\mathcal{G}}_{Bij} \cdot  k_{j} + \varepsilon_{i}\cdot \ddot{\mathcal{G}}_{Bij} \cdot \varepsilon_{j} \big]}\Big|_{\textrm{lin }\varepsilon_{1}\ldots  \varepsilon_{N}}\,.
	\label{eqMFScalarConst}
\end{align}
Expanding the Bern-Kosower exponent to multi-linear order will define polynomials $\mathcal{P}_{N}\dot{(\mathcal{G}}, \ddot{\mathcal{G}})$ in the background; the extension to the spin-orbit decomposition in spinor QED should be clear. 

Amongst the various applications of these formulae we mention worldline calculations of vacuum polarisation tensors \cite{Dittrich:2000wz, Schubert:2000yt}, to effects in a constant magnetic field \cite{Adler:1996cja, doi:10.1142/S0217732394002021}, and more recently to rederiving and extending the lately discovered one-particle-reducible contributions to the EHL starting at two-loop order in \cite{1PRScal, 1PRSpin, RedPaper, RedPRoc} which gave a streamlined proof that such contributions can be generated by differentiating lower order, irreducible diagrams.

\section{Worldline approach in plane wave backgrounds}
\label{secPlane}
In this section we follow \cite{NPlane} by incorporating plane wave backgrounds into the worldline framework. As indicated in the introduction we are motivated by models of intense laser environments and the efficiencies often gained with first quantised techniques. This work also extends \cite{AntonPlane} and clarifies important aspects of their analysis of helicity flip on the worldline. Indeed, since their work used a semi-classical analysis, which is usually exact only for Gaussian path integrals, it had been unclear how to expose this structure in a more general framework.

\subsection{Plane wave fields}
Our general setup is to define the plane wave background by a vector potential depending on a null vector, $n^{2} = 0$, so that $e\bar{\A}(x) = \arm(n\cdot x)$ with $\arm(-\infty) = 0$. The null vector defines light-cone coordinates $x^{+} := n\cdot x$ and $x^{-}, \mathbf{x}^{\perp}$ which imply the inner product 
\begin{equation}
	k \cdot x = -k^{+}x^{-} - k^{-}x^{+} + \mathbf{x}^{\perp}\cdot \mathbf{x}^{\perp}\,.
\end{equation}
We will assume light-front gauge, $n\cdot a = 0$, and later on it will be convenient to specify the gauge of external photons to satisfy $n\cdot \varepsilon_{i} = 0$ (valid for arbitrary momenta).  

The worldline representation of the one-loop $N$-photon amplitudes in this background is 
\begin{align}
	\hspace{-2em}\Gamma[k_{1}, \varepsilon_{1};\ldots;k_{N}, \varepsilon_{N}] = -\frac{1}{2}(-ie)^{N}&\int_{0}^{\infty}\frac{dT}{T}\e^{-im^{2}T} \oint_{PBC}\mathscr{D}x(\tau)\e^{i\int_{0}^{T}d\tau \,\big[\frac{\dot{x}^{2}}{4} - \dot{x}^{\mu}a_{\mu}(x^{+})\big]} \nonumber\\
	& \oint_{ABC}\mathscr{D}\psi(\tau)\, \e^{i\int_{0}^{T}d\tau \, \big[\frac{i}{2}\psi \cdot \dot{\psi} + i\psi^{\mu}f_{\mu\nu}(x^{+}) \psi^{\mu}\big]} \, \prod_{i=1}^{N} V^{\gamma}[k_{i}, \varepsilon_{i}] \,, \label{GammaWLa}
\end{align}
where $f_{\mu\nu}(x^{+}) = n_{\mu}a'_{\nu}(x^{+}) - a'_{\mu}(x^{+})n_{\nu}$. Inspecting the $x^{+}$ dependence of the plane wave background potential it is clear that the integrals over $x^{-}$ and $\mathbf{x}^{\perp}$ go through as before -- in particular integrating over $x_{0}^{+}$ and $\mathbf{x}_{0}^{\perp}$ gives momentum conservation in these directions.

The integral over $x^{+}$ seems to be more challenging due to the awkward $q(\tau)$ dependence of $a_{\mu}(x^{+}) \equiv a_{\mu}(x_{0}^{+} + q^{+}(\tau))$, but it turns out we can achieve a resummation that reveals the true Gaussian nature of the path integral. To see this, for the orbital coupling we expand
\begin{align}
\e^{-i\int_0^T d\tau   \,\dot q\cdot a(x_0^+ + q^+(\tau))}
=
1 &- i \int_0^T d\tau \, \dot q\cdot a(x_0^+ + q^+(\tau))
\nonumber\\&
+
\frac{(-i)^2}{2!} \int_0^Td\tau  \, \dot q\cdot a(x_0^+ + q^+)  \int_0^Td\tau'  \, \dot q'\cdot a(x_0^+ + q^{\prime +})
+
\cdots \,,
\end{align}
which will enter in the expectation value ($q_{i} \equiv q(\tau_{i})$ and $\psi_{i} \equiv \psi(\tau_{i})$ etc)
\begin{equation}
	\Big\langle \e^{-i\int_0^T d\tau   \,\dot q\cdot a(x_0^+ + q^+(\tau))} \, \prod_{i=1}^{N} V^{\gamma}[k_{i}, \varepsilon_{i}] \Big\rangle = \Big\langle \e^{-i\int_0^T d\tau   \,\dot q\cdot a(x_0^+ + q^+(\tau))} \, \prod_{i=1}^{N} \big[\varepsilon_{i}\cdot \dot{q}_{i} + \psi_{i}\cdot f_{i}\cdot \psi_{i}\big] \e^{i k_{i} \cdot (x_{0i} + q_{i})} \Big\rangle\,.
\end{equation}
Considering the possible Wick contractions of the field $q$, we see that for a non-vanishing result
\begin{itemize}
    \item We cannot contract a $q^{+}_{m} = n\cdot q_m$ with another such factor because $n^2=0$\,.
    \item We cannot contract a $q^{+}_{m} = n\cdot q_m$ with a $\dot{q}_l\cdot a$ because $n\cdot  a = 0$\,. 
\end{itemize}
This leaves the only remaining option, that each $n\cdot q_{m}$ must be contracted with the external photon vertex operators. As shown in \cite{NPlane} this implies a replacement in the argument of $a_{\mu}$
\begin{equation}
    q^{+}_{m} \longrightarrow n\cdot \sum_{i=1}^N\bigl[-i G_{mi}k_i + \dot G_{mi}\varepsilon_i\bigr]
    \label{eqResumWick}
\end{equation}
and the resummations that remove the path integration variable from the interaction exponential
\begin{equation}
  \hspace{-1em}  a_\mu(x_0^+ +  q_m^{+}) \longrightarrow a_\mu(\tau) \equiv a_\mu\Bigl(x_0^+ + \sum_{i=1}^N[-i G_{mi}k_i^{+} + \dot G_{mi}\varepsilon_i^{+}]\Bigr)\,, \qquad \e^{-i\int_0^T d\tau  \, \dot q\cdot a(x_0^+ +  q^{+}(\tau))}
\longrightarrow
\e^{-i\int_0^T d\tau   \dot q\cdot a(\tau)} \,,
	 \label{eqResuma}
\end{equation}
and render the integral in Gaussian form. This explains why the semi-classical solution in \cite{AntonPlane} obtained by expanding about the trajectory satisfying the classical equation of motion was exact.

\subsection{Master formula}
Having removed the path integration variable from the argument of $\arm$ we can compute the $N$-photon amplitudes with the simple procedure of completing the square in the exponent of 
\begin{equation}
	\int_{\textrm{SI}}\mathscr{D}q(\tau)\, \e^{i\int_{0}^{T}d\tau \big[\frac{\dot{q}^{2}}{4} -\dot{q}\cdot a(\tau) + \sum_{i=1}^{N} \big( k_{i} \cdot q_{i} - i\varepsilon_{i}\cdot \dot{q}_{i}\big)\big]}\,.
\end{equation}
The resulting effective action is most conveniently expressed via \textit{worldline averages} defined by\footnote{We direct the interested reader to \cite{AntonPlane} to see how to convert these to \textit{spacetime} averages.}
\begin{equation}
	\av f := \frac{1}{T}\int_{0}^{T}d\tau f(\tau)\,,
\end{equation}
in terms of which the analysis of \cite{NPlane} yields the Master Formula for scalar QED
\begin{align}
 \hspace{-1em}\Gamma_{\rm scal}[k_{1}, \varepsilon_{1};\ldots;k_{N}, \varepsilon_{N}] &=
(-ie)^N 
(2\pi)^3 
\delta\bigl(\sum_{i=1}^N k_i^1\bigr)
\delta\bigl(\sum_{i=1}^N k_i^2\bigr)
\delta\bigl(\sum_{i=1}^N k_i^+\bigr)
\totint dx_0^+ \e^{-i x_0^+ \sum_{i=1}^N k_i^-}
\nonumber\\
\hspace{-1em}&\hspace{-30pt}\times
\int_0^{\infty}
\frac{dT}{T}\,
{(4\pi iT)}^{-{D\over 2}}
\prod_{i=1}^N \int_0^Td\tau_i
\e^{
\frac{i}{2} \sum_{i,j=1}^N 
\bigl\lbrack   G_{ij} k_i\cdot k_j
-2i\dot G_{ij}\varepsilon_i\cdot k_j
+\ddot G_{ij}\varepsilon_i\cdot\varepsilon_j
\bigr\rbrack}
\nonumber\\
\hspace{-1em}&\hspace{-30pt}\times
\e^{-i\bigl(m^2+ \langle\langle a^2 \rangle\rangle - \langle\langle a \rangle\rangle^2\bigr)T
+2i\sum_{i=1}^N k_i \cdot 
\bigl(I(\tau_i)-\langle\langle I \rangle\rangle \bigr) 
+2\sum_{i=1}^N\bigl(a(\tau_i)-\langle\langle a\rangle\rangle \bigr) \cdot \varepsilon_i
}
\Bigl\vert_{\varepsilon_1\cdots \varepsilon_N}
\end{align}
where we also introduced the periodic integral of the background potential, 
\begin{equation}
	I_\mu(\tau) \equiv \int_0^\tau d\tau' \big(a_\mu(\tau') - \langle\langle a_\mu \rangle\rangle \big)\,.
\end{equation}
We can be more explicit by imposing the further gauge condition $n\cdot \varepsilon_{i} = 0$ which eliminates the polarisations from the argument of $a(\tau)$ in (\ref{eqResuma}) and facilitates the projection onto the multi-linear sector, realised with the help of new polynomials $\mathfrak{P}_{N}$ defined now by
\begin{equation}
	(-i)^{N} \mathfrak{P}_{N} \equiv {\rm e}^{\frac{i}{2}\sum_{i,j=1}^N\big(-2i\dot G_{ij}\varepsilon_i\cdot k_j+ \ddot G_{ij}\varepsilon_i\cdot \varepsilon_j\big)
+2 \sum_{i=1}^N\bigl(a(\tau_i)-\langle\langle a\rangle\rangle \bigr) \cdot \varepsilon_i
}
\big \vert_{\varepsilon_{1}\cdots\varepsilon_{N}}
\, .
	\label{eqPNa}
\end{equation}
With this we can write
\begin{align}
\label{eqGammaScala}
\hspace{-1.5em}\Gamma_{\rm scal}[k_{1}, \varepsilon_{1};\ldots;k_{N}, \varepsilon_{N}] &= (-e)^N
(2\pi)^3 
\delta\bigl(\sum_{i=1}^N k_i^1\bigr)
\delta\bigl(\sum_{i=1}^N k_i^2\bigr)
\delta\bigl(\sum_{i=1}^N k_i^+\bigr)
\!\!
\totint dx_0^+ \e^{-i x_0^+ \sum_{i=1}^N k_i^-}
\\
\hspace{-1.5em}&\hspace{-50pt} \times
\int_0^\infty
\frac{dT}{T}\,
{(4\pi iT)}^{-{D\over 2}}
\e^{-i\bigl(m^2+ \langle\langle a^2 \rangle\rangle - \langle\langle a \rangle\rangle^2\bigr)T}
\prod_{i=1}^N\int_0^Td\tau_i\,
\mathfrak{P}_{N} 
\,{\rm e}^{\frac{i}{2} \sum_{i,j=1}^N G_{ij}k_i\cdot k_j+2i\sum_{i=1}^N k_i \cdot 
\bigl(I(\tau_i)-\langle\langle I \rangle\rangle \bigr)}\,. \nonumber
\end{align}
Note that the proper time exponent has been completed to the effective mass in the background.

\subsubsection{Spinor QED}
For the spinor case, the spin-orbit decomposition still goes through based on the polynomials $\mathfrak{P}_{N}$ and a resummation of the spin interaction with the background:
\begin{equation}
	f_{\mu\nu}(x_{0}^{+} + q^{+}) \longrightarrow f_{\mu\nu}(\tau) \equiv n_{\mu}a'_{\nu}(\tau) - a'_{\mu}(\tau)n_{\nu}\,. 
	\label{eqfa}
\end{equation}
To deal with the Wick contractions for the spin interaction, then, we only need the Green function for the operator $\mathcal{O}_{\mu\nu} := \frac{\eta_{\mu\nu}}{2}\partial_{\tau} + ia'_{\mu}(\tau)n_{\nu} - in_{\mu}a'_{\nu}(\tau)$, inverted in \cite{NPlane} by
\begin{align}
 \hspace{-2em}  \langle \psi^\mu(\tau) \psi^\nu(\tau') \rangle &= \half \mathfrak G_F^{\mu \nu}(\tau,\tau')=\frac{1}{2}
\Bigl\lbrace \delta^{\mu \nu}\! +\! 2i n^\mu{\cal J}^\nu(\tau,\tau')\! +\! 2i {\cal J}^\mu(\tau',\tau)n^\nu
\! + \! 2\Bigl\lbrack {\cal J}^2(\tau,\tau')-\frac{T^2}{4} \langle\langle a'\rangle\rangle^2\Bigr\rbrack  
n^\mu n^\nu\Bigr\rbrace 
G_F(\tau,\tau')\,,
	\label{eqGFa}
\end{align}
in terms of further periodic functions of the background field 
\begin{align}
J_\mu(\tau) &\equiv \int_0^\tau d\tau' \Bigl( a'_\mu(\tau') - \langle\langle a'_\mu \rangle\rangle \Bigr)\, , \qquad {\cal J}_\mu(\tau,\tau') \equiv J_\mu(\tau)-J_\mu(\tau') - \frac{T}{2}\dot G (\tau,\tau')  \langle\langle a'_\mu \rangle\rangle \, .
	\label{eqDefJs}
\end{align}
Then, generalising (\ref{eqWS}) to
\begin{align}
\label{eqWSa}
\hspace{-1em}\mathfrak{W}(f_{i_1}\ldots,f_{i_S}) 
&\equiv  \Bigl\langle \psi_{i_1}\cdot f_{i_1} \cdot \psi_{i_1}
\cdots
\psi_{i_S}\cdot f_{i_S} \cdot \psi_{i_S}\Bigr\rangle_{\mathfrak{G}_F} \\
\hspace{-1em}&=\sum_{\rm partitions}
(-1)^{cy} \mathfrak{G}_F(i_1i_2\ldots i_{n_1})\mathfrak{G}_F(i_{n_1+1}\ldots i_{n_1+n_2})\cdots\mathfrak{G}_F(i_{n_1+\ldots + n_{cy-1}+1}\ldots i_{n_1+\ldots + n_{cy}})\, ,\nonumber
\end{align}
with $\mathfrak{G}_F(i_1i_2\dots i_{n}) := 2^{-\delta_{n2}}
\textrm{tr}(f_{i_1}\cdot{\mathfrak{G}}_{Fi_1i_2}\cdot f_{i_2}\cdot{\mathfrak{G}}_{Fi_2i_3}\cdots f_{i_n}\cdot{\mathfrak{G}}_{Fi_ni_1})$, and likewise (\ref{eqPNS}) to
\begin{align}
\hspace{-1.5em}(-i)^{N-S} \mathfrak{P}_{NS}^{\lbrace i_1i_2\ldots i_S\rbrace }  \equiv {\rm e}^{\frac{i}{2}\sum_{i,j=1}^N\big(-2i\dot G_{ij}\varepsilon_i\cdot k_j+\ddot G_{ij}\varepsilon_i\cdot \varepsilon_j\big)
+2 \sum_{i=1}^N\bigl(a(\tau_i)-\langle\langle a\rangle\rangle \bigr) \cdot \varepsilon_i
}
\Big \vert^{\varepsilon_{i_1}=0= \cdots 
 = 0=\varepsilon_{i_S}}_{\varepsilon_{i_{S+1}}\cdots\varepsilon_{i_N}} 
\, , 
\label{eqPNSa}
\end{align}
we may write explicitly the elements of the spin-obit decomposition in the background
\begin{align}
\label{GammaNSa}
\hspace{-2em}\Gamma_{NS}^{\lbrace i_1i_2\ldots i_S\rbrace}
&= -2 (-e)^N
(2\pi)^3 
\delta\bigl(\sum_{i=1}^N k_i^1\bigr)
\delta\bigl(\sum_{i=1}^N k_i^2\bigr)
\delta\bigl(\sum_{i=1}^N k_i^+\bigr)
\totint dx_0^+ \e^{-i x_0^+ \sum_{i=1}^N k_i^-}
\\
\hspace{-2em}&\hspace{-50pt} \times
\int_0^\infty
\frac{dT}{T}\,
{(4\pi i T)}^{-{D\over 2}}
\e^{-i\bigl(m^2+ \langle\langle a^2 \rangle\rangle - \langle\langle a \rangle\rangle^2\bigr)T}
\prod_{i=1}^N\int_0^Td\tau_i\,
\mathfrak{W} (f_{i_1},\ldots,f_{i_S}) 
\mathfrak{P}_{NS}^{\lbrace i_1i_2\ldots i_S\rbrace} 
{\rm e}^{\frac{i}{2} \sum_{i,j=1}^N G_{ij}k_i\cdot k_j+2i\sum_{i=1}^N k_i \cdot 
\bigl(I(\tau_i)-\langle\langle I \rangle\rangle \bigr)}
\, . \nonumber
\end{align}
In \cite{NPlane} the Master Formulae for both scalar and spinor QED are expanded to order $N = 2$ photons, first working completely off-shell, before specialising to the on-shell case and finally confirming the amplitude for helicity flip studied using the semi-classical approach in \cite{AntonPlane}.
\section{Conclusion}
\label{secConc}
This contribution has reviewed one-loop photon scattering amplitudes in the Worldline Formalism of quantum field theory, starting in vacuum and then including a constant and later plane wave backgrounds. We have presented Master Formulae for the $N$-photon amplitudes, expected to streamline calculations, especially for higher $N$, since they combine various Feynman diagrams and the propagators in the loop into compact integral representations that avoids the need to fix a specific ordering or integrate explicitly over Volkov functions. It also has the advantage of being a unifying description for the scalar and spinor theories which appear more similar in this formalism than in the standard approach.

In ongoing work we are analysing the application of the Master Formulae for the $N = 3$ (photon-splitting) and $N=4$ (light-by-light scattering) amplitudes in the plane wave background. The formalism is also being extended to the open-line case to describe the scalar and spinor propagators in the plane wave background, building upon \cite{Ahmad:2016vvw, fppaper1, fppaper2}.

\ack

The authors thank Anton Ilderton and Antonino di Piazza for helpful correspondence. JPE acknowledges funding through a CIC--U.M.S.N.H project.


\section*{References}

\bibliographystyle{iopart-num}
\bibliography{bibPlane.bib}

\end{document}